\documentclass[twocolumns,letter]{rmaa}

\usepackage{gensymb}

\title{CCD PHOTOMETRY OF THE GLOBULAR CLUSTER $NGC$ $6093$\altaffilmark{1}}

\author{A. Ruelas-Mayorga\altaffilmark{2}, L. J. S\'anchez\altaffilmark{2}, C. A. Bernal-Herrera \altaffilmark{2}, A. Nigoche-Netro\altaffilmark{3}, J. Echevarr\'{\i}a\altaffilmark{2} and A. M. Garc\'{\i}a\altaffilmark{2}}

\altaffiltext{1}{Based upon observations acquired at the Observatorio Astron\'omico Nacional on the Sierra San Pedro M\'artir (OAN-SPM), Baja California, M\'exico.}
\altaffiltext{2}{Instituto de
Astronom\'{\i}a, Universidad Nacional Aut\'onoma de M\'exico,
M\'exico D.F., M\'exico.}
\altaffiltext{3}{Instituto de Astronom{\'\i}a y Meteorolog{\'\i}a, Universidad
     de Guadalajara, Guadalajara, Jal. 44130, M\'exico.}

\fulladdresses{
 \item A. Ruelas-Mayorga, L.~J.~S\'anchez, C. A. Bernal-Herrera, J. Echevarr\'{\i}a, A. M. Garc\'{\i}a: Instituto de Astronom\'{\i}a, Universidad
                                                                                                                 Nacional Aut\'onoma de M\'exico, Apartado Postal 70-264, Cd.
                                                                                                                 Universitaria 04510, M\'exico D.F., M\'exico.
                                                                                                                 (\email{rarm,leonardo@astro.unam.mx,cbernal@cenagas.gob.mx,jer,agarcia@astro.unam.mx
                                                                                                                 }).
\item A. Nigoche-Netro: Instituto de Astronom{\'\i}a y Meteorolog{\'\i}a, Universidad
                                                                          de Guadalajara, Guadalajara, Jal. 44130, M\'exico. (\email{anigoche@gmail.com}).}

\shortauthor{Ruelas-Mayorga et al.}
\shorttitle{CCD Photometry of NGC 6093}

\abstract{In this paper we present photometric CCD observations of the globular cluster NGC 6093 (M80) in filters $B$, $V$, $R$ and $I$.
We produce the colour-magnitude diagrams for this object and obtain values for its metallicity $[Fe/H]$, reddening $E(V-B)$, $E(V-I)$
and distance modulus $(m-M)_{0}$.}

\resumen{En este art{\'\i}culo presentamos observaciones fotom\'etricas del c\'umulo globular NGC 6093 (M80) en los filtros
$B$, $V$, $R$ e $I$. Producimos los diagramas color-magnitud para este objeto y obtenemos los valores de su metalicidad
$[Fe/H]$, enrrojecimiento $E(V-B)$, $E(V-I)$, y m\'odulo de distancia $(m-M)_{0}$.}


\begin{document}

\maketitle

\section{INTRODUCTION}

\label{sec:intro}

The Globular Clusters in our Galaxy are very rich stellar clusters which present spherical symmetry,
and may be found in regions closer to the Galactic Centre as well as in remote regions of the
Galactic Halo. They contain hundreds of thousands of stars within a radius $20-50$ $pc$, having typical central densities
between $10^{2}$ and $10^{4}$  $stars/pc^{3}$. Globular clusters live for a long time and are dynamically very stable, they
are remnants of a primordial star formation epoch and are present in all galaxies that are relatively large.  They
may be considered proper galactic subsystems \citep{ruelas}.

A large number of Milky Way globular clusters have been known for a long time (see the Messier ($1771$) catalogue). At present
we estimate their number to be approximately 150, although this figure does not include very low surface brightness objects
or those close to the plane of the Galaxy, \citet{monella} and \citet{globulardata}.

One of the reasons why globular clusters are fundamental in the study of Galactic Structure, is the fact that they are
very luminous systems and, therefore, may be observed at very large distances. In general the light that comes from these
objects originates in stars slightly cooler than our Sun.

Some characteristics of Globular Clusters are:

\begin{itemize}

\item Morphology. In general they are slightly elliptical in shape. The average ratio between the minor and major axes of
the apparent ellipse that they project on the sky is $b/a=0$.$93$, with only $5\%$ of them more elongated than $b/a=0$.$8$.
Besides, they appear to be exclusively stellar system in which no presence of gas or dust is detected \citep{galactic}.

\item The value of their integrated absolute magnitude $(M_{V})_{0}$ is usually found in the interval $-5>$($M_{V}$)$_{0}\gtrsim -10$
where the maximum of the distribution is found at $(M_{V})_{0}\approx -7$.$5$ and with a FWHM of $\sim \pm 1$ mag. The radial distribution
of stars varies between clusters, and there are some that present a strong central concentration \citep{HARRIS2010}.

\item Their intrinsic colour takes values in the interval  $0$.$4\lesssim (B-V)_{0}\lesssim 0$.$8$ with a maximum at
$(B-V)_{0}\approx 0$.$57$ \citep{MIHALAS,HARRIS2010}.

\item It is common to find clusters with metallic abundances in the interval $-2$.$2\lesssim \left[ Fe/H\right]\lesssim 0$.$0$. There
are clusters with a large metallic deficiency, usually located in the Galactic Halo, up to those with abundances similar to that
of the Sun \citep{MIHALAS}.

\end{itemize}

The study of Globular Clusters is important because they may be used for a variety of astronomical purposes, such as: determination of
the Galactic Centre position, studies of the evolution of low mass and low metallicity stars, chemical evolution of galactic systems
and also as indicators of the galactic gravitational potential.

\begin{figure}[!htbp]
\begin{center}
\includegraphics[width=8.0cm,height=8.0cm]{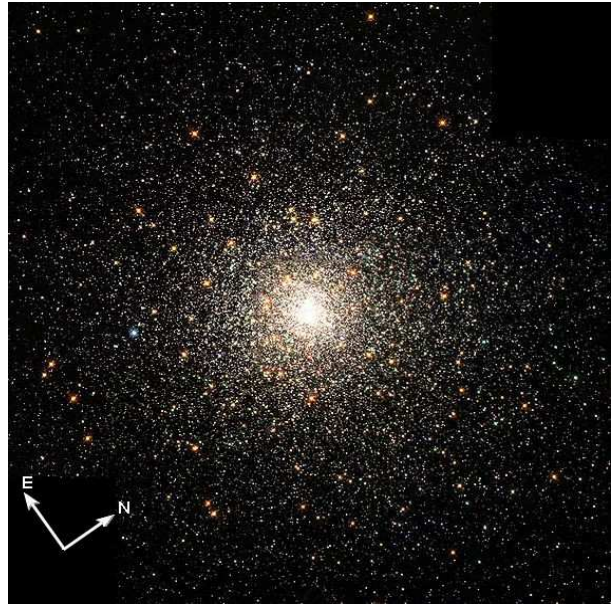}

\end{center}
\caption{The Globular Cluster NGC\ 6093
(M80). The image is $3\arcmin$ vertically (Taken from:
http://hubblesite.org/newscenter/).} \label{fig:m80}
\end{figure}

The globular cluster which we study in this paper is NGC 6093, also known as M80, and is located on the Scorpius constellation ($AR (2000:$ $16^{h}17^{m}02$.$41^{s}$,
$DEC (2000:$ $-22\degree 58^{\arcmin }33.9^{\arcsec }$), it contains several hundred thousands stars and it is one of the densest globular
clusters in the Galaxy \citep{m80} (see Figure \ref{fig:m80}). Its most important properties are listed in Table \ref{tab:m80tabla}
\citep{globulardata}.

\begin{table*}[!htbp]
\caption{{}}
\label{tab:m80tabla}\centering
\begin{tabular}{llc}
\hline\hline
\multicolumn{3}{c}{Data for the Globular Cluster NGC 6093 (M80)} \\ \hline
\multicolumn{2}{l} {Right Ascension (2000)} & $16^{h}17^{m}02$.$41^{s}$ \\
\multicolumn{2}{l}{Declination (2000)} & $-22\degree 58^{\arcmin }33.9^{\arcsec }$ \\
\multicolumn{2}{l}{Galactic Longitude} & $352$.$67$ \\
\multicolumn{2}{l}{Galactic Latitude} & $19$.$46$ \\
\multicolumn{2}{l}{Distance to the Sun(kpc)} & $10.0$ \\
\multicolumn{2}{l}{Distance to the Galactic Centre (kpc)} & $3$.$8$ \\
\multicolumn{2}{l}{Reddening $E(B-V)$} & $0$.$18$ \\
\multicolumn{2}{l}{Horizontal Branch Magnitude (in $V$)} & $16$.$1$ \\
\multicolumn{2}{l}{Distance Modulus ($m-M$) (in $V$)} & $15$.$56$ \\
\multicolumn{2}{l}{Integrated $V$ Magnitude} & $7$.$33$ \\
\multicolumn{2}{l}{Absolute Visual Magnitude} & $-8$.$23$ \\
& $U-B$ & $0$.$21$ \\
Integrated Colour Indices & $B-V$ & $0$.$84$ \\
(no reddening correction) & $V-R$ & $0$.$56$ \\
& $V-I$ & $1$.$11$ \\
\multicolumn{2}{l}{Metallicity $[Fe/H]$} & $-1$.$75$ \\
\multicolumn{2}{l}{Integrated Spectral Type} & $F6$ \\
\multicolumn{2}{l}{Heliocentric Radial Velocity (km$/$s)} & $8$.$1$ \\
\multicolumn{2}{l}{Central Concentration} & $0$.$79$ \\
\multicolumn{2}{l}{Ellipticity} & $0$.$00$ \\
\multicolumn{2}{l}{Core Radius (arcmin)} & $0.15$ \\
\multicolumn{2}{l}{Mean Mass Radius (arcmin)} & $2$.$11$ \\
\multicolumn{2}{l}{Tidal Radius (arcmin)} & $13$.$28$ \\
\multicolumn{2}{l}{Central Surface Brightness (in $V$) (magnitudes$/$%
arcsec$^2$)} & $15$.$11$ \\
\multicolumn{2}{l}{Logarithm of the luminous density at the centre ($L_{\odot
}/$pc$^{3}$)} & $4$.$79$ \\ \hline
\end{tabular}
\end{table*}

This object is of interest because there are few ground-based photometric studies of its stars in the Johnson-Cousins system. Up to now the only available photometric studies have been published by: Harris \& Racine (1974), Brocato et al. (1998), Alcaino et al. (1998), Rosenberg et al. (2000) have obtained $V$ and $I$ photometry for 52 nearby globular clusters included NGC 6093, and there are also Hubble Space Telescope (HST) observations of this cluster (see Ferraro et al. 1999a, and Piotto et al. 2002).

In addition, it has a large number of \textit{Blue
Stragglers} in its nucleus (see Ferraro et al. 1999b). As is well know, these stars are bright and blue stars that appear close to the main sequence of the HR diagram
and appear to be younger and more massive than the rest of the cluster stars. The presence of these objects has been interpreted as
an indication of a rather large rate of collisions in the nucleus of this globular cluster.

It presents a Horizontal Branch of very hot stars (\textit{blue tail}), which extends over a large interval of effective
temperature. It has been conjectured that there exists a correlation between stellar populations of this sort and the high density
of some globular clusters (Buonanno et al. 1985).

In this paper we intend to determine, from our photometric data, the Metallicity $[Fe/H]$, the reddening $E(B-V)$, $E(V-I)$ and the distance to $NGC6093$.

In Section 2 we present the Observations and the reduction process, Section 3 presents the colour-magnitude diagrams which were derived for several combinations of filters, Section 4 presents the Analysis used to calculate the Metallicity, the Reddening and the Distance Modulus of this cluster
and, finally, Section 5 presents our Conclusions.

\qquad

\section{THE OBSERVATIONS}
\label{sec:obs}

We obtained the observations at the Observatorio Astron\'omico
Nacional in San Pedro M\'artir (OAN-SPM), Baja California during 2006, March 20-23 and 2007, March 14.

We utilised two different CCD cameras attached to the 1.5\,m telescope. The characteristics of these detectors are presented
in Table \ref{tab:detectordata}.

\begin{table*}[!htbp]
\caption{ }
\label{tab:detectordata}\centering
\begin{tabular}{ccc}
\hline\hline
\multicolumn{3}{c}{Characteristics of the Detectors Used in the Observations} \\ \hline
Characteristic & Thomson & Site1 \\ \hline
\multicolumn{1}{l}{Size (pixels)} & $2048\times 2048$ & $1024\times 1024$ \\
\multicolumn{1}{l}{Pixel Size ($\mu m$)} & $14\times 14$            & $24\times 24$ \\
\multicolumn{1}{l}{Quantum Efficiency}   & Maximum 65\% at 5000 \AA &  \\
\multicolumn{1}{l}{Reading noise ($e^{-}$) (gain } &  &  \\
mode $4$ binning $2\times 2$) & $5$.$3$ & $5$.$5$ (Direct Imaging) \\
\multicolumn{1}{l}{Dark Current ($e^{-}$/pix/h)} & $1$.$0$ a $-95$.$2$ºC & $7$.$2$ approximately at $-80$ºC \\
\multicolumn{1}{l}{Well Depth ($e^{-}$)} & $1$.$23\times 10^{5}$
(MPP Mode) &  \\
\multicolumn{1}{l}{Bias Level (gain } &  &  \\
mode $4$ binning $2\times 2$) & $384$ & $547$ (Direct Imaging) \\
\multicolumn{1}{l}{Gain ($e^{-}$) (Mode  $4$)} & $0$.$51$ & $1$.$27$ \\
\multicolumn{1}{l}{A/D Converter} & $16$ bits &  \\
\multicolumn{1}{l}{Linearity} &  $99\%$ &  $99$.$5\%$ \\
\multicolumn{1}{l}{Plate Scale ($\arcsec$/pixel)} & $0$.$147$
& $0$.$252$ \\ \hline
\end{tabular}
\end{table*}

As stated above, the observations were collected during two observing runs, during which, we observed in a standard way
obtaining bias and flat field frames in each filter, plus obtaining multiple observations of four standard stars regions; namely Rubin 149, 152, and PG 1323 and 1525.
Plus 10 frames per filter (B, V, R, I) of the central region of NGC 6093. We present in Table \ref{tab:bitacora} an example of a log of observations for the
night March 20-21, 2006 to show the sequence followed during the observations.

\begin{table*}[!htbp]
\caption{ }
\label{tab:bitacora}\centering%
\begin{tabular}{ccccccc}
\hline\hline
NUMBER       & UNIVERSAL            &   OBJECT       &   TYPE             &        AIR MASS        &         EXPOSURE TIME          &    FILTER       \\
OF FRAMES    & TIME                 &                &                    &                        &         PER FRAME (s)          &                  \\
\hline
  10         & 11:24:52-11:28:46    &   PG1528       &   Standard Star    &         1.10           &             10                 &      B           \\
  10         & 11:34:35-11:41:30    &   PG1528       &   Standard Star    &         1.11           &             30                 &      R            \\
  10         & 11:44:20-11:51:15    &   PG1528       &   Standard Star    &         1.11           &             30                 &      I            \\
  10         & 11:55:18-12:02:12    &   PG1528       &   Standard Star    &         1.12           &             30                 &      V            \\
             &                      &                &                    &                        &                                &                   \\
  10         & 12:21:01-12:25:41    &  NGC 6093      &   Globular Cluster &         1.71           &             15                 &      I            \\
  10         & 12:31:40-12:36:20    &  NGC 6093      &   Globular Cluster &         1.72           &             15                 &      R            \\
  10         & 12:50:45-13:02:10    &  NGC 6093      &   Globular Cluster &         1.75           &             60                 &      V            \\
  10         & 13:16:15-13:28:45    &  NGC 6093      &   Globular Cluster &         1.78           &             60                 &      B            \\
  \hline

\end{tabular}
\end{table*}

The observed frames for the standard
stars in the 2006 observing season were not reliable because the filter wheel suffered an unspecified slippage during the observations,
and it resulted impossible for us to identify the observed standard frames with a definite filter, moreover in some occasions the filter wheel may have been stuck half way, so the observations of the standard stars may have suffered from severe vignetting. This problem was corrected before we obtained the observations for the globular cluster.

\begin{figure}[!htbp]
\begin{center}
\includegraphics[width=8.0cm,height=8.0cm]{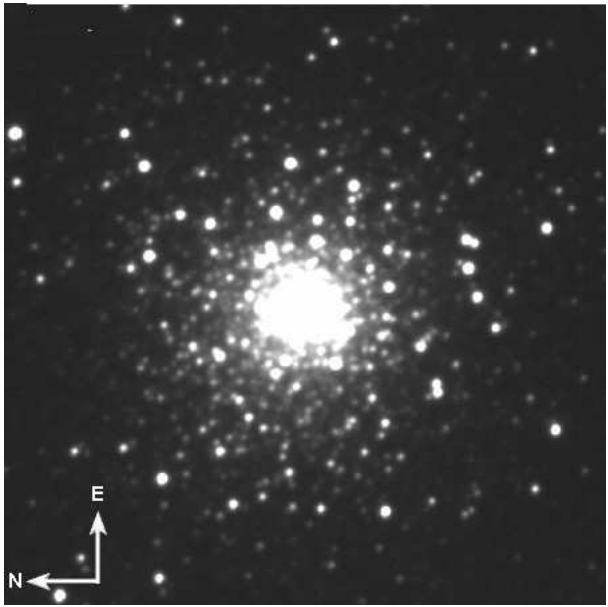}

\end{center}
\caption{Image of the  central zone of the Globular Cluster NGC 6093 in the V filter after the process of preparation.} \label{fig:imagenreducida2006b}
\end{figure}

The reduction of the data was done in a standard way, that is, removing dead and hot pixels produced by cosmic rays,
bias subtraction, and flat field correction. This reduction process was achieved using the general purpose software: Image Reduction and Analysis Facility (IRAF). Once the fields
have been bias and flat-field corrected we proceed to utilise the routine DAOPHOT to obtain the photometry of the
many starts present in the field using the Point Spread Function (PSF) technique (Stetson, 1992). As an example of an image
on which we apply the DAOPHOT technique see Figure \ref{fig:imagenreducida2006b}.

\subsection{Reduction to the Standard System}

In order to express the magnitude of the stars in the globular cluster in a standard system, we performed aperture photometry
of stars in four Landolt Standard Regions (Landolt, 1992), namely; Rubin 149, 152, and PG 1323 and 1525. This photometry was carried out using
the IDL data reduction programme: ´ATV.pro´, and measuring the brightness of each star in 10 or fewer apertures of different size.  The number of counts measured in each ring was corrected for background brightness, which was measured
in an empty sky-region near the standard star. During the measuring process, ´ATV.pro´ defines two regions, a central circular region and an external annular region which is used to measure the background intensity. One has to be careful that in this external annulus there are no stars at all, otherwise this sky measurement is useless. These corrected measurements were fitted to a curve of growth expressed by:

\begin{equation}
y(x)=A\left( 1-e^{-\frac{x}{D}}\right)
\end{equation}

in which $y$ represents the number of counts at a distance $x$ from the centre of the star, and D is a characteristic separation from the centre. $A$ is the asymptotic value which $y$ obtains when $x$ tends to infinity, and which is the value adopted for the
number of counts for the measured star in question.

The reduction of the standard stars in the Landolt Regions permitted us to establish that the process of atmospheric
extinction not only depends on the value of the airmass $(X)$ but it also presents a slight variation with the intrinsic colour
of the different stars used in the process. We, therefore, proposed a set of transformation equations from the observed to the
intrinsic values of the magnitudes in the standard system of the following form:

\begin{equation}
B_{int}-B_{obs}=A_B X+K_B\left(B-V\right)_{int}+C_B
\label{eq:eqforB}
\end{equation}

\begin{equation}
V_{int}-V_{obs}=A_V X+K_V\left(B-V\right)_{int}+C_V
\label{eq:eqforV}
\end{equation}

\begin{equation}
R_{int}-R_{obs}=A_R X+K_R\left(R-I\right)_{int}+C_R
\label{eq:eqforR}
\end{equation}

\begin{equation}
I_{int}-I_{obs}=A_I X+K_I\left(R-I\right)_{int}+C_I
\label{eq:eqforI}
\end{equation}

The values of the constants $A$, $K$ and $C$ for each filter are found by a least square process, which consists in using the observed magnitude for the standard stars and their intrinsic values from the Landolt catalogue in the transformation equations. A least square fit produces the values for constants $A$, $K$ and $C$ for each filter.

The values of the intrinsic magnitudes for the stars in the globular cluster
are calculated from the equations that result when solving simultaneously for $B_{int}$ and $V_{int}$ from equations \ref{eq:eqforB}
and \ref{eq:eqforV} and for $R_{int}$ and $I_{int}$ from equations \ref{eq:eqforR} and \ref{eq:eqforI}. The easiest way to solve equations
\ref{eq:eqforB}, and \ref{eq:eqforV}, and \ref{eq:eqforR} and \ref{eq:eqforI} simultaneously is to find the values of $(B-V)_{int}$ and
$(R-I)_{int}$ by subtracting from equation \ref{eq:eqforB} equation \ref{eq:eqforV} and subtracting from equation \ref{eq:eqforR} equation \ref{eq:eqforI}.

The resulting solutions for the intrinsic colours look as follows:

\begin{equation}
(B-V)_{int}=\frac{(B-V)_{obs}+(A_{B}-A_{V})X+(C_{B}-C_{V})}{1-(K_{B}-K_{V})}
\end{equation}

\begin{equation}
(R-I)_{int}=\frac{(R-I)_{obs}+(A_{R}-A_{I})X+(C_{R}-C_{I})}{1-(K_{R}-K_{I})}
\end{equation}

By substitution of the values of the intrinsic colours in equations \ref{eq:eqforB}, \ref{eq:eqforV}, \ref{eq:eqforR} and \ref{eq:eqforI}, the values of the intrinsic magnitudes are then easily found.

The values for the constants $A$, $K$, and $C$ for all the filters, and their errors are shown in Table \ref{tab:coeficiente}.

\begin{center}
\begin{table}[!htbp]
\caption{ }
\label{tab:coeficiente}\centering
\begin{tabular}{cccc}
\hline\hline
\multicolumn{4}{c}{Fit Coefficients for each Filter}
 \\ \hline
$Filter$           & $A$                 &  $K$                & $C$                  \\ \hline
$B$                & $-0$.$22 \pm 0.01$  &  $0$.$14 \pm 0.01$  & $24$.$40 \pm 0.05$    \\
$V$                & $-0$.$09 \pm 0.01$  &  $0$.$02 \pm 0.01$  & $24$.$60 \pm 0.05$    \\
$R$                & $-0$.$03 \pm 0.02$  &  $0$.$12 \pm 0.02$  & $24$.$63 \pm 0.05$    \\
$I$                & $-0$.$14 \pm 0.02$  &  $0$.$27 \pm 0.02$  & $24$.$35 \pm 0.10$    \\ \hline
\end{tabular}%
\end{table}
\end{center}

\begin{figure*}[!htbp]
\begin{center}
\includegraphics[width=15.0cm,height=14.0cm]{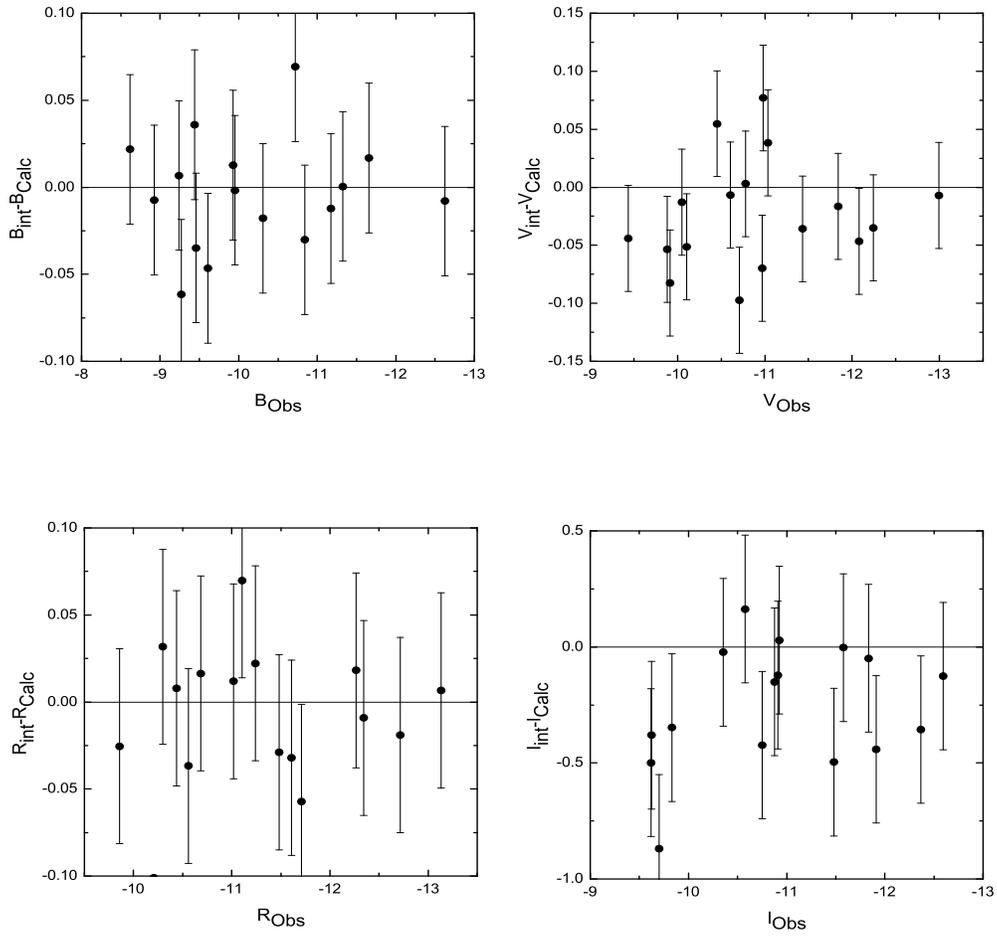}

\end{center}
\caption{Residuals for all filters for standard stars.} \label{fig:panel_residuals}
\end{figure*}

In Figure \ref{fig:panel_residuals} we present the typical residuals
obtained for standard stars to equations \ref{eq:eqforB}, \ref{eq:eqforV}, \ref{eq:eqforR} and \ref{eq:eqforI} using the coefficients
in Table \ref{tab:coeficiente}. The standards used were PG1323 stars 0, A to C, Rubin 149 stars 0, A to F and Rubin 152 stars 0, A to C and E and F. Region PG 1525 was not used for having very large residuals.

\subsection{Photometry of the Cluster Stars}

The calculation of the different observed magnitudes of the stars in the cluster NGC6093
was obtained by a standard application of the DAOPHOT subroutine present in IRAF to all the
observed frames in each filter, after having the frames bias-subtracted and flat fielded.

\begin{figure*}[!htbp]
\begin{center}
\includegraphics[width=15.0cm,height=14.0cm]{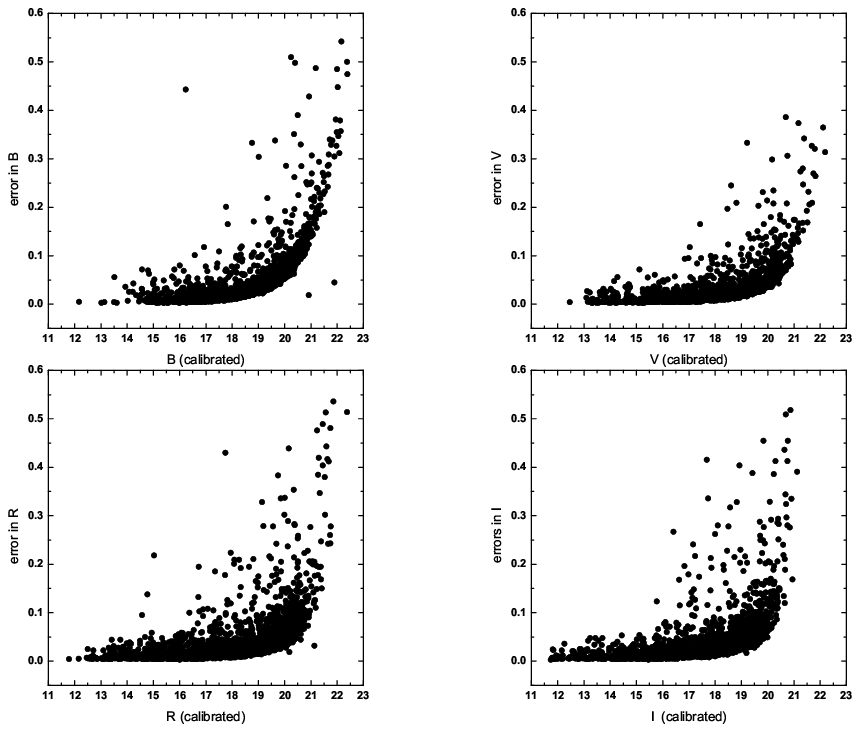}
\end{center}
\caption{Errors for each filter versus magnitude.} \label{fig:panelerrores}
\end{figure*}

The magnitude catalogues produced by the subroutine ALLSTAR for the 2007 observations were combined
to obtained the observed $(B-V)_{obs}$ and $(R-I)_{obs}$ colours for the stars in the globular cluster, so that the application of equations
\ref{eq:eqforB}, \ref{eq:eqforV}, \ref{eq:eqforR} and \ref{eq:eqforI} could be made in a straightforward manner. In the Figure \ref{fig:panelerrores} we present the errors of the measured magnitudes as functions of the calibrated magnitudes, as expected, the errors increase as the magnitudes become fainter.

As mentioned before (see Section\ref{sec:obs}) the observations of the standard stars in the 2006 observing run were not reliable, so, in order to calibrate the 2006 cluster stars observations, we identified common stars to the 2006 and 2007 observing runs in each filter. Then we took
the value of the observed magnitude in the 2006 run $m_{obs}$ and the intrinsic value of the magnitude from the 2007 run $m_{int}$, and
we constructed a relation of the following form:

\begin{equation}
m_{int}=Am_{obs}+b
\label{eq:eq2006to2007}
\end{equation}

Calculation of the constants $A$ and $b$ allow the calculation of the intrinsic magnitudes of the stars present in the 2006 frames
for all the filters.

\begin{figure*}[!htbp]
\begin{center}
\includegraphics[width=14.0cm,height=15.0cm]{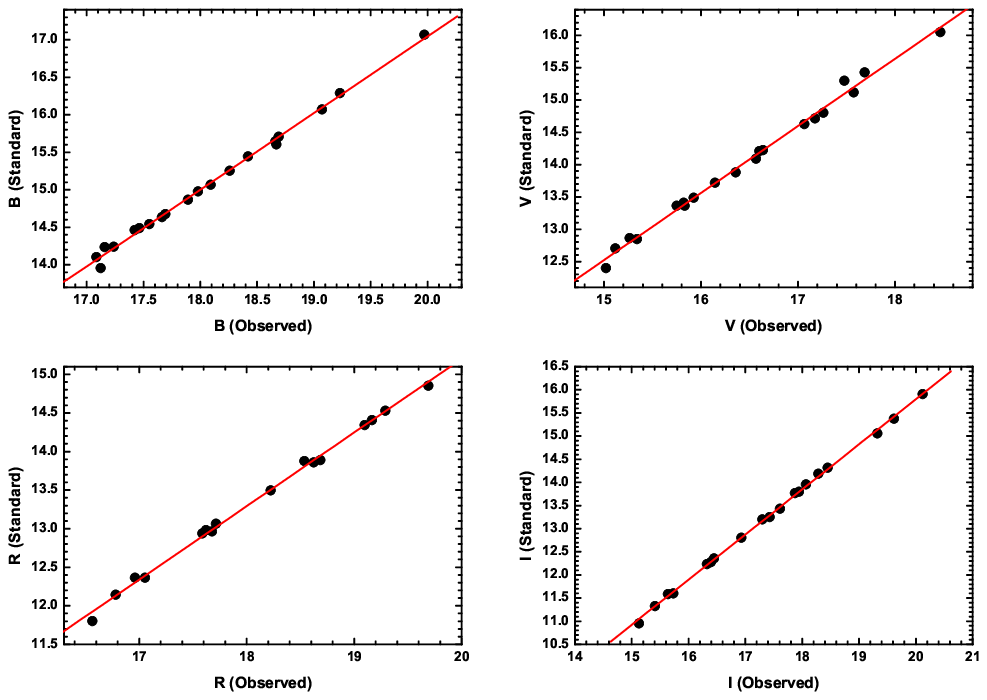}
\end{center}
\caption{Transformation for the four filters used between the observed and intrinsic magnitudes for the 2006 observing run.} \label{fig:TRANSFORMATIONS}
\end{figure*}

The values of the transformation coefficients for each filter are given in  Table \ref{tab:coefs2006} and the fits to
the common stars in each filter are shown in Figure \ref{fig:TRANSFORMATIONS}.

%
%
\begin{center}
\begin{table}[!htbp]
\caption{ }
\label{tab:coefs2006}\centering%
\begin{tabular}{lcc}
\hline\hline
\multicolumn{3}{c}{Transformation Coefficients between} \\
\multicolumn{3}{c}{observing runs in 2006 and 2007} \\ \hline
& $A$ & $b$ \\ \hline
$B$ & $1$.$015\pm 0$.$014$ & $-3$.$265\pm 0$.$258$ \\
$V$ & $1$.$040\pm 0$.$019$ & $-3$.$073\pm 0$.$316$ \\
$R$ & $0$.$967\pm 0$.$013$ & $-4$.$107\pm 0$.$232$ \\
$I$ & $0$.$975\pm 0$.$006$ & $-3$.$702\pm 0$.$111$ \\ \hline
\end{tabular}%
\end{table}
\end{center}

\section{Colour-Magnitude Diagram}
\label{sec:HRdiagram}

\begin{figure*}[!htbp]
\begin{center}
\includegraphics[width=16.0cm,height=20.0cm]{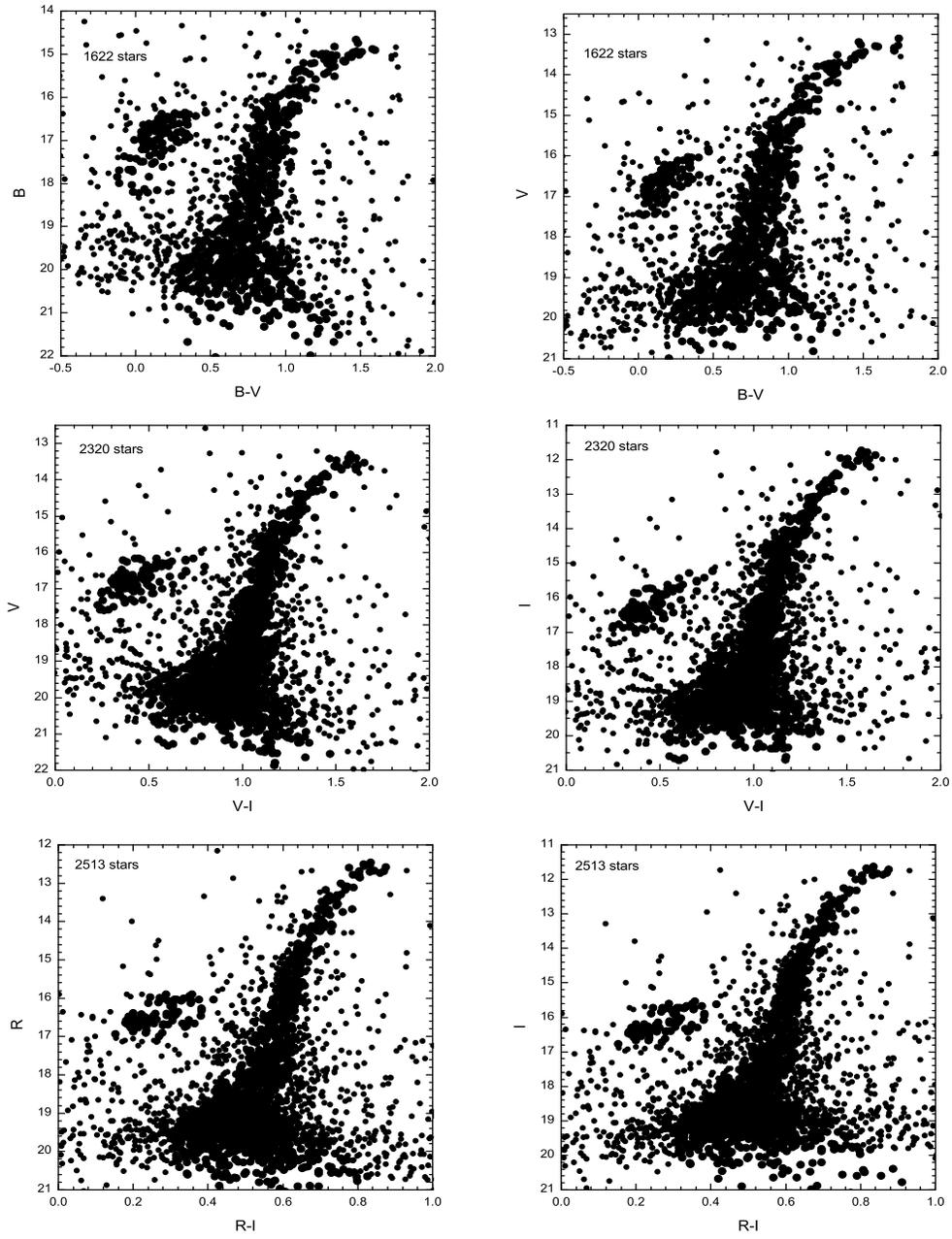}
\end{center}
\caption{Colour-Magnitude diagrams for NGC 6093.} \label{fig:PANELCM}
\end{figure*}

Once the stars in all the images are translated to a common reference frame, their calculated intrinsic magnitudes
may be used to obtain their intrinsic $B-V$, $R-I$ and $V-I$ colours. A number of colour-magnitude diagrams are
produced and are shown in Figure \ref{fig:PANELCM}.

In all the CMDs it is clear that the magnitude limits of the present observations
is around 20 mag, where the photometric errors begin to be very large and
measures are spread. This prevents to see a clear sub-giant branch and
of course the Turn Off (TO) point. An estimation from the CMD by Brocato
et al. 1998 and Rosenberg et al 2000 leads to a TO magnitude level
around 19.6 mag. The TO V-magnitude estimated in Alcaino et al. 1998
is brighter ($19.16 \pm 0.05$ mag). It is also clear the presence of the Giant Branch (GB) and the Horizontal Branch (HB), the top of the Main Sequence
may be guessed at the fainter magnitudes presented ($ \sim +20$ for all magnitudes).

\section{Analysis}
\label{sec:fiducial}

In this section we derive values for the metallicity $([Fe/H])$, the reddening $(E(B-V)$, $E(V-I)$ and the distance modulus $((m-M)_0)$
of the globular cluster NGC6093.

\subsection{Fiducial Line}
\label{subsec:fiducial}

\begin{figure}[!htbp]
\begin{center}
\includegraphics[width=8.0cm,height=8.0cm]{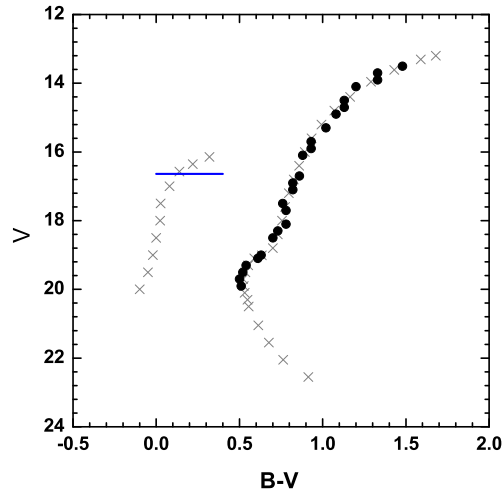}
\end{center}
\caption{Fiducial lines obtained in this work (full circles) and that of Brocato et al. (1998) (crosses). The level of the HB and its colour extent is shown on this figure.} \label{fig:fiducials}
\end{figure}

The first step in this analysis is to distinguish between those stars that belong to the cluster and those that are field stars, which
found their way into the colour-magnitude diagrams just because they overlap the cluster's field in the sky. In order to achieve this, we
obtain the colour distribution of the stars in a given colour-magnitude diagram at a fixed magnitude level $(M_{fid})$, and find the value of the colour $(CI_{fid})$
at which the number of objects in that bin is maximum.
This procedure produces the coordinates of a point on the colour-magnitude diagram $(CI_{fid},M_{fid})$ and the value of the standard deviation $(\sigma)$
of the colour distribution at that magnitude level.
Repeating this process over the entire range of magnitudes covered by the colour-magnitude diagram produces a set of points which, when joined,
give a smooth curve which is denominated in the literature as the {\it fiducial line}. The {\it fiducial line} define the geometric
locus where the Main Sequence and the Giant Branch of a cluster are located on the colour-magnitude diagram. In this paper we shall assume
that a star that is further than $1.5 \sigma$  from the {\it fiducial line} at its corresponding magnitude level does not belong to the cluster.
In Figure \ref{fig:fiducials} we see the fiducial line obtained in this work (full circles) compared with that obtained by
Brocato et al. (1998) (crosses).

\begin{figure*}[!htbp]
\begin{center}
\includegraphics[width=16.0cm,height=8.0cm]{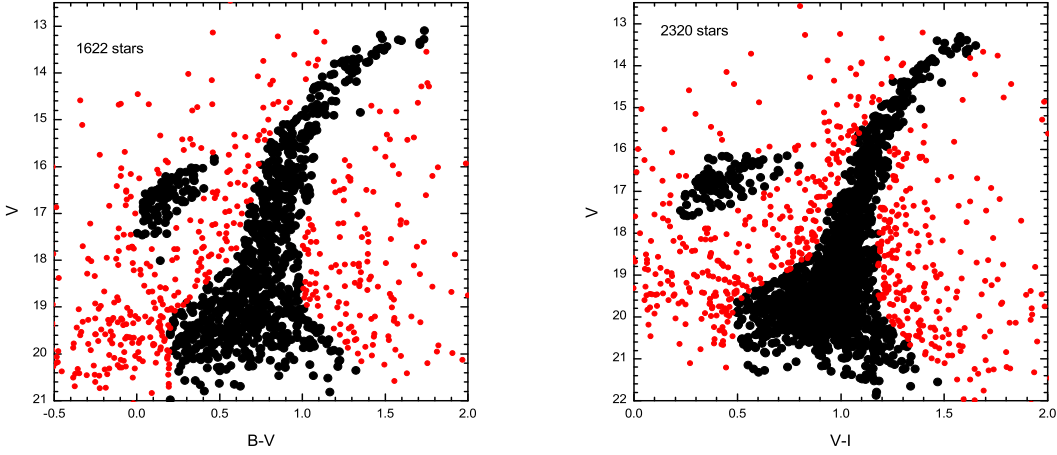}
\end{center}
\caption{$V$ vs $B-V$ and $V$ vs $V-I$ colour-magnitude diagrams. The stars in red are those eliminated from the analysis for having a large
probability of not being cluster members.} \label{fig:dos_referee}
\end{figure*}

The subsequent analyses that we shall present in subsections \ref{subsec:metred}, and \ref{subsec:modulus}
are carried out using only those stars that belong to the cluster according to the criterion established above. For the analysis presented in
Subsection \ref{subsec:metred}, we used the $V$ vs $B-V$ and the $V$ vs $V-I$ colour-magnitude diagrams, the stars that appear in red in Figure \ref{fig:dos_referee}
are those that have a large probability of not being cluster member according to the criterion presented above.

\subsection{Metallicity and Reddening}
\label{subsec:metred}

Metallicity and reddening for a globular cluster may be determined simultaneously by direct application of the Sarajedini method
(Sarajedini, 1994 and Sarajedini \& Layden, 1997), which takes into consideration the shape of the Red Giant Branch (RGB),
the observed magnitude $(V_{HB})$ of the Horizontal Branch (HB), the intrinsic $(B-V)$ colour of the RGB at the level of the HB; this value will be denoted herein as: $(B-V)_{0,g}$, and the difference in observed $V$ magnitude between the HB and the RGB at $(B-V)_{int}=1.2$
denoted by $\Delta V_{1.2}=V_{HB\ Obs}-V_{RGB\ Obs\ at \  1.2 \  int}$, where $V_{RGB\ Obs\ at \  1.2 \  int}$ means the observed $V$ magnitude of the RGB at an intrinsic $(B-V)$ colour equal to $1.2$. $\Delta V_{1.2} $ results in an essentially positive quantity.

The Sarajedini relations in the $B$ vs $B-V$ plane are:

\begin{equation}
\left[  Fe/H\right]  =-5.763+5.236(B-V)_{0,g}
\label{eq:eqmetBV1}
\end{equation}

and

\begin{equation}
\left[  Fe/H\right]  =0.3492-0.8618\Delta V_{1.2}
\label{eq:eqmetBV2}
\end{equation}

If $(B-V)_{g}$ is the observed colour of the RGB at the level of the HB, then  $E\left(
B-V\right) =(B-V)_{g}-(B-V)_{0,g}$. We can, therefore, rewrite  (\ref{eq:eqmetBV1}) as:

\begin{equation}
E(B-V)   =-0.19098\left[  Fe/H\right]  +\\
 (B-V)_{g}-1.10065.
\label{eq:mari}
\end{equation}

Now, we fit a second degree polynomial to the points of the RGB on the observed $V$ vs $(B-V)$ plane. The equation chosen for the second degree polynomial is as follows:

\begin{equation}
(B-V)=a_{0}+a_{1}V+a_{2}V^{2}
\label{eq:ajusRGR}
\end{equation}

Equation (\ref{eq:ajusRGR}) is a general fit to the observed points of the RGB. We shall now apply this equation to a particular point, that for which the intrinsic $(B-V)$ colour is equal to $1.2$. At this point we may write the following two equations (\ref{eq:auxiliares1} and \ref{eq:auxiliares2}):

\begin{equation}
B-V=1.2+E(B-V)
\label{eq:auxiliares1}
\end{equation}

\begin{equation}
V=V_{HB}-\Delta V_{1.2}.
\label{eq:auxiliares2}
\end{equation}

where $B-V$ represents the observed colour of the RGB which when dereddened will take an intrinsic value of $1.2$, $V$ is the intrinsic value of the
average magnitude of the HB and $V_{HB}$ is the observed average magnitude value of the HB.

Then, substituting (\ref{eq:auxiliares1}) and (\ref{eq:auxiliares2}) in (\ref{eq:ajusRGR}) we get:

\begin{equation}
E(B-V)=b_{0}+b_{1}\Delta V_{1.2}+b_{2}(\Delta V_{1.2})^{2}
\label{eq:enroj}
\end{equation}%

where

\begin{equation}
b_{0}=a_{0}+a_{1}V_{HB}+a_{2}V_{HB}^{2}-1.2
\end{equation}

\begin{equation}
b_{1}=-a_{1}-2a_{2}V_{HB}
\end{equation}

\begin{equation}
b_{2}=a_{2}.
\end{equation}

Substituting expression (\ref{eq:enroj}) in
(\ref{eq:mari}), we finally obtain that

\begin{equation}
k_{0}+k_{1}\left[  Fe/H\right]  +k_{2}\left[  Fe/H\right]  ^{2}=0
\label{eq:metalicidad}
\end{equation}

being

\begin{equation}
k_{0}=b_{0}+0.4052b_{1}+0.1642b_{2}+1.1006-(B-V)_{g}
\end{equation}

\begin{equation}
k_{1}=0.191-1.1604b_{1}-0.9404b_{2}
\end{equation}

\begin{equation}
k_{2}=1.3465b_{2}.
\end{equation}

Therefore, knowing the values of $V_{HB}$, $(B-V)_{g}$ and the coefficients
$a_{i}$ from the fit to the RGB, it is possible to solve equations (\ref{eq:metalicidad}) for the metallicity $[Fe/H]$. The value of
$\Delta V_{1.2}$ may be calculated from equation (\ref{eq:eqmetBV2}) and then the colour excess $E(B-V)$ from equation (\ref{eq:enroj}).

For the HR diagram in the $V$ vs. $\left( B-V\right) $ plane, the nominal values and the
fit coefficients are:

\begin{eqnarray}
a_{0} =9.920\pm 0.906,   \\
a_{1} =-0.994\pm 0.114,   \\
a_{2} =0.027\pm 0.004, \\
\left( B-V\right) _{g} =0.83,  \\
V_{HB} =16.6.
\end{eqnarray}

Sarajedini also produced a calibration in the $V$ vs. $(V-I)$ space. Which is:

\begin{equation}
p_{1}+p_{2}[Fe/H]+p_{3}[Fe/H]^{2}=0
\end{equation}

and

\begin{equation}
E(V-I)=(V-I)_{g}-0.1034\left[  Fe/H\right]  -1.100,
\end{equation}

and

\begin{equation}
\Delta V _{1.2}=-1.068\left[  Fe/H\right]+0.2782,
\end{equation}

being

\begin{equation}
p_{1}=b_{0}+0.2782b_{1}+0.0774b_{2}+1.100-(V-I)_{g}
\end{equation}

\begin{equation}
p_{2}=0.1034-1.068b_{1}-0.5942b_{2}
\end{equation}

\begin{equation}
p_{3}=1.141b_{2}
\end{equation}

defining the $b_{i}$
coefficients as before, but considering the new RGB. In this case:

\begin{eqnarray}
a_{0} =7.030\pm 0.419,   \\
a_{1} =-0.638\pm 0.052,   \\
a_{2} =0.017\pm 0.002, \\
\left( V-I\right) _{g} =1.05,   \\
V_{HB} =16.6.
\end{eqnarray}

\bigskip In order to calculate $E(B-V)$ from $E(V-I)$ we use the relation given by
Dean et al. (1978) with $\left( B-V\right) _{0}=1.2$ :

\begin{equation}
\frac{E(V-I)}{E(B-V)}=1.250\left[  1+0.06(B-V)_{0}+0.014E(B-V)\right]
\label{eq:dean}
\end{equation}

For the globular cluster studied in this paper, the metallicity and the reddening (colour excess) obtained using this method using
both techniques, are shown in Table \ref{tab:resultadosfin}. The colour excess values marked
with an asterisk were obtained using Dean's relation(eq. \ref{eq:dean}).

We adopt as final values: $[Fe/H]=-1.84\pm 0.20$, which has the smallest error,
and $\ E(B-V)=0.15\pm 0.03$, that corresponds to the mean value of the values obtained,
given that the associated uncertainty is the same. Using Dean's formula we get
$E(V-I)=0.20 \pm 0.04$. Note that these values are consistent with
those found in the literature\footnote{%
Brocato, et al. (1998) calculate the metallicity $\left[ Fe/H\right] $
for $NGC6093$ using correlations with $(B-V)_{0,g}\,$, $\Delta V_{1.4}$
 and with the $S$ parameter. The mean weighted value obtained in that paper
applying the first method is $\left[ Fe/H\right] =-1.94\pm 0.13$, which
is very close to the results presented here.} (See Table \ref{tab:comparacion}).

\begin{table*}[!htbp]
\caption{{}}
\label{tab:resultadosfin}\centering%
\begin{tabular}{lccc}
\hline\hline
\multicolumn{4}{c}{Results} \\ \hline
HR Diagram & $\left[ Fe/H\right] $ & $E(B-V)$ & $E(V-I)$ \\ \hline
$V$ vs $B-V$ & $-1.84\pm 0.20$ & $0.14\pm 0.03$  & $0.19\pm 0.02$*\\
$V$ vs $V-I$ & $-1.95\pm 0.26$ & $0.16\pm 0.03$* & $0.22\pm 0.03$\\
\multicolumn{4}{l} {* Using Dean's relation eq. (\ref{eq:dean})}  \\
\hline
\end{tabular}%
\end{table*}

\begin{table*}[!htbp]
\caption{{}}
\label{tab:comparacion}\centering%
\begin{tabular}{lcc}
\hline\hline
\multicolumn{3}{c}{Metallicity and Reddening values for NGC 6093} \\ \hline
Reference & $[Fe/H]$ & $E(B-V)$ \\ \hline
Racine (1973) &  & $0$.$17$ \\
Harris \& Racine (1974) &  & $0$.$16$ \\
Bica \& Pastoriza (1983) & $-1$.$55$ &  \\
Zinn (1985) & $-1$.$68$ &  \\
Brodie \& Hanes (1986) & $-1$.$57$ &  \\
Reed et al. (1988) &  & $0$.$18$ \\
Suntzeff et al. (1991) & $-1$.$64$ &  \\
Brocato (1998) & $-1$.$71$ &  \\
Alcaino et al (1998) & $-1$.$7$ &  \\
This work (2012) & $-1$.$84\pm 0$.$20$ & $0$.$15\pm 0$.$003$ \\ \hline
\end{tabular}%
\end{table*}

\subsection{Distance Modulus}
\label{subsec:modulus}

In order to calculate the distance modulus to the globular cluster we study in this paper, we
shall use the following assumption: the average absolute magnitude of its HB is equal to that of
the RR-Lyrae stars, for a justification of this statement see, for example, Christy (1966),  Demarque \& McClure (1977) and
Saio  (1977).

The number of variable stars found in NGC 6093 is, according to Clement et al. (2001) equal to 8, out of which 6 are of the RR-Lyrae type,
1 is a Cepheid or an RV Taurus star and the other one has not had its period determined, so it has not been assigned to any specific type of
variable star. Four of the RR-Lyraes in this cluster pulsate in the fundamental mode, and 2 in the first overtone, with average periods of 0.651 and 0.366 days respectively. The Horizontal Branch Ratio (HBR) for this cluster is equal to 0.93 indicating that its HB is, for the most part, composed of
blue stars.

It has become customary in the astronomical literature to assume, for the RR-Lyrae stars, a linear relation between absolute magnitude and metallicity
of the form $M=a+b[Fe/H]$, the problem is then reduced to finding the value of the constants $a$ and $b$. The most popular methods utilised for calibrating
the values of the $a$ and $b$ constants are: i) through statistical parallaxes, ii) the BBW moving atmosphere method, and iii) main sequence fitting
(Sandage,~A. \& Tammann,~G.~A., 2006).

In the following paragraphs we shall use different absolute magnitude-metallicity relations
from the literature
for RR-Lyraes, which combined with the metallicity and the apparent magnitude for the
HB of NGC 6093 given in this paper, will permit us to determine the value of its distance modulus.

\begin{table*}[!htbp]
\caption{ } \label{tab:distancemodulusNGC6093} \centering
\setlength{\tabcolsep}{0.5\tabcolsep}
\begin{tabular}{ccc}
\hline\hline
\multicolumn{3}{c}{DISTANCE MODULUS FOR NGC 6093}\\\hline

   From the Calibration given in     &     ${\langle \mathrm{M_V}\rangle}              $  &   $\rm{(m-M)_0}$  \\
                                     &                                                    & $m_{HB}-\langle \mathrm{M_V}\rangle-3.1E(B-V)$ \\\hline
                                     &                                                    &                     \\
Tsujimoto, Miyamoto \& Yoshii (1998) &           $0.54 \pm 0.52$                          &   $15.64 \pm 0.84$  \\
Wan, Mao \& Ji (1980)                &           $0.86 \pm 0.10$                          &   $15.32 \pm 0.42$  \\
Skillen et al. (1993)                &           $0.65 \pm 0.19$                          &   $15.53 \pm 0.51$  \\
McNamara (1997)                      &           $0.45 \pm 0.14$                          &   $15.73 \pm 0.46$  \\
Fernley (1993)                       &           $0.49 \pm 0.04$                          &   $15.69 \pm 0.36$  \\
Arellano-Ferro et al. (2008)         &           $0.52 \pm 0.11$                          &   $15.66 \pm 0.43$  \\
\hline
\end{tabular}
\end{table*}

Tsujimoto, Miyamoto, \& Yoshii (1998) have analysed data for 125 {\it Hipparcos}
RR Lyraes in the metallicity range $-2.49 <$ [Fe/H] $< 0.07$ using the maximum likelihood
technique proposed by Smith (1988). This technique allows simultaneous correction of the Malmquist and Lutz-Keller biases, allowing a full
use of negative and low-accuracy parallaxes. They derive the following relation:
$${\langle \mathrm{M_V\rangle}_{RR}} = (0.59 \pm 0.37) +
(0.20 \pm  0.63)({\rm [Fe/H]} +1.60) \, .$$
\noindent Given that $\mathrm{[Fe/H]_{NGC\ 6093}}=-1.84 \pm 0.20$ is contained within the studied
metallicity interval, applying this relation to the cluster studied in this paper produces
$${\langle \mathrm{M_V}\rangle}_{\rm NGC\ 6093} = 0.54 \pm 0.52 \, .$$

A compilation of statistical parallaxes of field RR-Lyrae stars has been presented by Wan, Mao, \& Ji (1980) in a Catalogue of the
Shanghai Observatory. This compilation is summarised in
Table~3 of Reid (1999). There
is a value for ${\langle \mathrm{M_V}\rangle}_{\rm RR} = 0.83
\pm 0.23$ for $\langle{\rm [Fe/H]}\rangle$ values around $-$0.75
and another ${\langle \mathrm{M_V}\rangle}_{\rm RR} = 0.85 \pm
0.15$ for $\langle{\rm [Fe/H]}\rangle$ values around $-$1.56. A
linear extrapolation to the metallicity of NGC 6093 (${\rm [Fe/H]}=
-1.84 \pm 0.20$) produces a value for
$${\langle \mathrm{M_V}\rangle}_{\rm NGC\ 6093} = 0.86 \pm 0.10 \, .$$

Even though the error for this determination is relatively small, the extrapolation process
makes this determination less reliable than the previous one.

Using a combination of the infrared flux and the Baade-Wesselink analysis methods Fernley et al. (1989, 1990a, 1990b), and Skillen et al. (1989, 1993) study 21 RR-Lyrae variable stars and obtain a mean relation for their absolute magnitude expressed as follows:
$$ {\langle \mathrm{M_V}\rangle}_{\rm RR} = (0.21\pm0.05){\rm [Fe/H]} + (1.04\pm0.10) \, ,$$
\noindent which for the metallicity value of our globular cluster produces a result of
$${\langle \mathrm{M_V}\rangle}_{\rm NGC\ 6093} = 0.65 \pm 0.19 \, .$$

McNamara (1997) has reanalysed these same 29 stars making use of
more recent Kurucz model atmospheres and derives a steeper, more
luminous calibration given as follows:
$$ {\langle \mathrm{M_V}\rangle}_{\rm RR} = (0.29\pm0.05){\rm [Fe/H]} + (0.98\pm0.04) \, ,$$

The RR-Lyraes studied in the McNamara paper belong to
a metallicity interval from approximately -2.2 to 0.0. The metallicity
value for our globular cluster (-1.84) lies within this interval, making it reasonable to apply this relation to this cluster.
\noindent The value we obtain from this relation is:
$${\langle \mathrm{M_V}\rangle}_{\rm NGC\ 6093} = 0.45 \pm 0.14 \, .$$
\noindent Fernley (1993) uses his near-IR Sandage Period-shift
Effect (SPSE) and a theoretical pulsation relation to derive the following relation:
$${\langle \mathrm{M_{V}}\rangle}_{\rm RR} = 0.19 {\rm [Fe/H]} + 0.84 \, ,$$
\noindent which applied to our cluster gives
$${\langle \mathrm{M_V}\rangle}_{\rm NGC6093} = 0.49 \pm 0.04 \, .$$

Arellano-Ferro et al. (2008a, b) using the technique of Fourier
decomposition for the light curves of RR-Lyraes in several
globular clusters derive the following relation:

$${\langle \mathrm{M_{V}}\rangle}_{\rm RR} = +(0.18 \pm 0.03) {\rm [Fe/H]} + (0.85 \pm 0.05) \, .$$

This relation was obtained for a set of globular clusters contained within the metallicity interval
$-2.2 < \rm{[Fe/H]} < -1.2$ making it appropriate for the metallicity value (-1.84) we find for NGC 6093 in this paper.

Applying this relation to our cluster we find

$${\langle \mathrm{M_V}\rangle}_{\rm NGC\ 6093} = 0.52 \pm 0.11 \, .$$

There are many different empirical and theoretical determinations
of the ${\langle \mathrm{M_V}\rangle}-{\rm [Fe/H]}$ relation for
RR-Lyrae stars, for ample discussions see Chaboyer (1999), Cacciari
\& Clementini (2003) and Sandage \& Tamman (2006). Determining
which one is the most appropriate for NGC 6093 is beyond the scope of
this paper, so we have decided to consider all of them for the calculation of the distance modulus of the
globular cluster studied in this paper.

From the data presented in
this paper we determine an apparent V
magnitude for the HB of NGC 6093 of $16.64 \pm 0.32$, which combined with the values for
the absolute magnitudes of the RR-Lyrae stars and the assumption that the HB and the RR-Lyraes have the same absolute magnitude yields
the distance modulus values presented in Table~\ref{tab:distancemodulusNGC6093}.

A weighted average (by the inverse square of the errors) of these values results in an average distance modulus for
NGC 6093 of $15.59 \pm 0.50$ $(13100^{+3400}_{-2700} \ pc)$. Which represents a $ \sim \pm 27 \%$ error in distance.
For the most part
the error in the distance modulus comes from the errors in the absolute magnitude versus
metallicity relations (see Table~\ref{tab:distancemodulusNGC6093}, column 2), and not from
the errors in our photometry.

Benedict et al. (2002), using the HST parallax for the prototype
RR-Lyrae star, determine an absolute magnitude for this star of
$M_v = 0.61 \pm 0.10$. If we assume that the HB of NGC 6093 has this
value for its absolute magnitude then we obtain a distance modulus
of:
$$\rm{(m-M)_0}=(16.64 \pm 0.32)-(0.61 \pm 0.10)$$
$$-3.1 \times (0.14 \pm 0.03)=15.60 \pm 0.51$$

which agrees, within the errors, with previous determinations. This value
of the distance modulus produces a distance of $13200^{+3500}_{-2800}$ pc
to NGC 6093 with an error of $ \sim \pm 27 \%$.

We adopt as our best determination for the NGC 6093 distance modulus the value
obtained with the Fernley(1993) calibration ($15.69 \pm 0.36$),
due to the fact that this calibration presents the smallest errors.

\section{Conclusions}
\label{sec:conclusions}

In this paper we have presented CCD photometry of the globular cluster NGC 6093 in the filters B, V, R and I, which
permitted us to produce six colour-magnitude diagrams (see Section \ref{sec:HRdiagram}) containing on the average $\sim 2000$
stars each. The observations were collected during two observing seasons in 2006 and 2007 as explained in Section \ref{sec:obs}.

In order to eliminate fore and background stars we produced for each colour-magnitude diagram a fiducial line and only kept those
stars that lied close to this line (see Section \ref{sec:fiducial}). Further analysis was performed only on the
stars that, according to the criterion established in Section \ref{sec:fiducial}, were truly cluster members.

Using the Sarajedini method (Sarajedini, 1994 and Sarajedini \& Layden, 1997) we calculated simultaneously the values of
the metallicity and the reddening of this globular cluster. The values of the colour excess in $(B-V)$ and in $(V-I)$ given here correspond to the average values given in Table \ref{tab:resultadosfin} resulting the following figures:

$$ [Fe/H]=-1.84 \pm 0.20$$

$$E(B-V)=0.15 \pm 0.03$$

$$E(V-I)=0.20 \pm 0.04$$

These values are consistent with those found in the literature as shown in Table \ref{tab:comparacion}.

We determined the distance modulus to NGC 6093 assuming that the average absolute magnitude of the stars in the
HB of the cluster is the same as that found for the RR-Lyrae stars. Using several Metallicity-Absolute Magnitude relations for
 RR-Lyraes (see Subsection \ref{subsec:modulus}), we found, for each, a distance modulus for NGC 6093, but decided to adopt the value with the smallest error. The distance modulus value given in this paper is equal to $15.69 \pm 0.36$. and corresponds to the relation given in Fernley (1993).

In a forthcoming paper we shall use these data to determine this cluster's Helium abundance and age by making use of
several sets of theoretical isochrones.

\acknowledgments

We would like to thank the comments of an anonymous referee, they improved greatly the presentation of this paper. We thank the support provided by Instituto de Astronom\'{\i}a at Universidad Nacional Aut\'onoma de M\'exico (UNAM), and Instituto de Astronom{\'\i}a y Meteorolog{\'\i}a, Universidad de Guadalajara (UG).
 We would like to thank Juan Carlos Yustis for his help with the figures presented in this paper. We also thank Direcci\'on General de Asuntos del Personal Acad\'emico, DGAPA at UNAM for financial support under projects number PAPIIT IN103813 and PAPIIT IN111713.

\end{document}